\documentclass[10pt,final,doublecolumn]{IEEEtran}
\usepackage{amsmath}
\usepackage{latexsym}
\usepackage{graphicx}
\usepackage{bbding}
\usepackage{indentfirst}
\usepackage{algorithm,algorithmic}
\usepackage{setspace}
\usepackage{float}
\usepackage{epstopdf}
\usepackage{amssymb}
\usepackage{amsfonts}
\usepackage{enumerate}
\usepackage{multicol}
\usepackage{color}
\usepackage{slashbox}
\usepackage{bm}
\usepackage{amssymb}
\usepackage{stfloats}
\usepackage{epstopdf}
\usepackage{threeparttable}
\usepackage{epstopdf}
\usepackage{threeparttable}
\usepackage{subfigure}
\usepackage{makecell}

\usepackage{cite}
\usepackage[table]{xcolor}

\usepackage[linewidth=0.5pt]{mdframed}
\IEEEoverridecommandlockouts
\allowdisplaybreaks[4]

\begin{document}
\title{6DMA Enhanced Wireless Network with Flexible Antenna Position and Rotation: Opportunities and Challenges
}











%






\author{{Xiaodan Shao,~\IEEEmembership{Member,~IEEE} and Rui Zhang, \IEEEmembership{Fellow, IEEE}}

	 \thanks{X. Shao is with Friedrich-Alexander-University Erlangen-Nuremberg.}

\thanks{R. Zhang is with The Chinese University of Hong Kong, Shenzhen,
Shenzhen Research Institute of Big Data, and the National University of
Singapore.
}
}
\maketitle

\IEEEpeerreviewmaketitle

\begin{abstract}
6DMA (six-dimensional movable antenna) is a new and revolutionizing technology
that fully exploits the wireless channel spatial variation at the transmitter/receiver by flexibly adjusting the three-dimensional (3D) positions and 3D rotations of distributed antennas/antenna surfaces (arrays).
In this article, we provide an overview of 6DMA for unveiling its great potential in wireless networks, including its motivation and
competitive advantages over existing technologies, system/channel modeling, and practical implementation. In particular, we present a variety of 6DMA-enabled  performance enhancement in terms of array gain, spatial multiplexing, interference suppression, and geometric gain. Furthermore, we illustrate the main applications of 6DMA in wireless communication and sensing, and elaborate their design
challenges as well as promising solutions. Finally, numerical results are provided to demonstrate the significant capacity improvement of 6DMA-aided communication in wireless network.
\end{abstract}

\section{Introduction}
To meet the demand for increasingly higher data rates with a steadily growing population of Internet-of-things (IoT) devices in the forthcoming six-generation (6G) wireless network, the current trend of multi-antenna or multiple-input multiple-output (MIMO) technologies is to equip the base station (BS) and wireless terminal with drastically more antennas.
For example, by deploying an even larger number
of antennas than the existing massive MIMO at the BS, the
extremely large-scale MIMO can achieve significantly higher spatial
degrees of freedom (DoFs) to enhance the wireless system communication/sensing performance \cite{MIMOD}. On the other hand, cell-free massive MIMO has been recently developed to provide users with seamless and uniform performance
coverage in the network by deploying more distributed antennas/BSs \cite{cellfreeq}. Moreover, the recent emergence of intelligent reflecting
surface (IRS)/reconfigurable intelligent surface (RIS) \cite{proc} has opened a new avenue for exploiting passive MIMO systems for wireless network performance enhancement with low-cost and power-efficient reconfigurable elements.

The multi-antenna technologies mentioned above can be regarded as
fixed-position MIMO, which employs fixed-position
antennas (FPAs) with their positions fixed
once deployed. As a result, their performance improvement by using more FPAs comes at
the cost of increasing hardware cost and energy consumption, in both fully-digital and hybrid analog/digital MIMO systems. In addition, given the
number of antennas at the transmitter/receiver, the
wireless network with FPAs cannot allocate its antenna resources
flexibly based on the user channel spatial distribution, beyond the traditional adaptive MIMO processing. Thus, FPAs may not be able to fulfill the high performance and efficiency expectation
of future wireless networks such as 6G, and research on finding innovative, high-performance, and yet cost-effective MIMO technologies for future wireless networks is still imperative.

Recently, to fully exploit the wireless channel spatial variation at the BS/wireless terminal, the six-dimensional movable antenna (6DMA) technology has been
proposed as a novel solution to improve the performance of wireless networks cost-effectively \cite{shao20246d, 6dma_dis}. Equipped with distributed antennas/antenna surfaces (arrays) that can be independently adjusted in three-dimensional (3D) position and 3D rotation, 6DMA-enabled transceivers are endowed with more spatial DoFs to adapt to the wireless channel spatial distribution dynamically and thereby enhance the network performance.
In practice, the positions and rotations of 6DMA surfaces can be adjusted continuously \cite{shao20246d} or in discrete levels \cite{6dma_dis}, depending on the surface movement mechanism, and optimized with or without prior knowledge of the channel spatial distribution in the network \cite{shao20246d, 6dma_dis}.

As compared to other existing MIMO technologies which are closely related to 6DMA, namely, Remote Electrical Tilt (RET) antennas \cite{ret} and fluid antenna system (FAS), also
called movable antenna (MA) \cite{9388928, zhulet,qingm}, the main differences as well as competitive advantages of 6DMA are summarized in Table I. Specifically, FAS/MA adjusts the positions of active antennas on a finite 2D surface only, without considering the antennas' rotation; while the RET antennas
offer limited movement flexibility by adjusting the vertical tilt angle of each sector antenna array as a whole. In contrast, 6DMA can more flexibly adjust the 3D positions and 3D rotations of antennas/antenna surfaces (arrays) via real-time controllable motors, which is thus capable of more effectively adapting to the user channel spatial distribution.
Moreover, in existing works on FAS/MA\cite{9388928,zhulet,qingm}, individual antennas usually serve as the basic movable units to maximally exploit the small-scale channel variation for mitigating the effect of deep fading. Since small-scale channel fading usually changes rapidly over a short period (on the order of milliseconds) or a short distance (on the order of signal wavelength), FAS/MA requires frequent antenna movement which incurs high implementation/hardware cost and time overhead. In contrast, 6DMA needs to be adjusted in position and/or rotation much less frequently, as its main performance gain stems from the adaptive allocation of antenna resources and spatial DoFs based on the channel spatial distribution, which varies slowly in practice (on an hourly, daily, or even longer-period basis). Although 6DMA may incur a moderately higher hardware/energy cost than existing RET antennas with antenna tilt adjustment only, it is expected to achieve much higher performance gains than RET via its more flexible antenna positioning and rotation.
\newcommand{\tabincell}[2]{\begin{tabular}
{@{}#1@{}}#2\end{tabular}}
\begin{table*}[!t]
\vspace{-0.6cm}
\small
\caption{Comparison of different antenna architectures}
\vspace{-0.3cm}
\centering
\begin{tabular}{>{\columncolor{blue!15}}c >{\columncolor{black! 10}}c >{\columncolor{blue!15}}c >{\columncolor{black! 10}} c >{\columncolor{blue!15}}c >{\columncolor{black! 10}}c >{\columncolor{blue!15}}c>{\columncolor{blue!15}}c}
\bfseries \tabincell{c}{Antenna \\Architecture} &\bfseries Movability & \bfseries {\tabincell{c}{Movable Unit}}
 & \bfseries \tabincell{c}{Movement Frequency} &\bfseries \tabincell{c}{Performance \\Gain}
&\bfseries Hardware &\bfseries Cost
\\
\Xhline{1pt}
6DMA \cite{shao20246d, 6dma_dis} & \tabincell{c}{High \\(position and rotation)}   & \tabincell{c}{Single antenna\\/antenna surface\\ (array) }  & \tabincell{c}{Low or pre-configured \\(based on channel \\spatial distribution) }  & Very high& Motor& Moderate\\
\Xhline{0.5pt}
\tabincell{c}{FAS/MA \\\cite{9388928,zhulet,qingm}}  &\tabincell{c}{Medium \\ (position only)}  & Single antenna  &\tabincell{c}{High (based on small-scale\\ channel fading) }& High& \tabincell{c}{Pixel antenna\\/motor}&High\\
\Xhline{0.5pt}
{RET \cite{ret}} & \tabincell{c}{Low \\(tilt angle only)}  & \tabincell{c}{Sector antenna\\ array} &\tabincell{c}{Low\\ (based on user distribution)}& High& {Motor}& Low
\end{tabular}
\label{table}
\vspace{-0.59cm}
\end{table*}

Despite its promising advantages and potential performance gains, the research on 6DMA-empowered wireless networks is still in its infancy and there are new challenges in implementing this technology and optimizing its performance in wireless networks. This thus
motivates this article to provide an overview of
6DMA, including its system/channel modeling, main performance enhancement, practical implementation, and promising applications. In particular, the main challenges and their potential solutions
for designing and implementing 6DMA-aided wireless networks are elaborated to inspire future research. Numerical results are also provided to validate the effectiveness of 6DMA for communication throughput enhancement.

\section{6DMA Modeling}
In this section, we first provide the general system model for 6DMA and discuss its practical movement constraints. Then, we introduce the 6DMA channel model.
\subsection{6DMA System Model}
\begin{figure*}[t!]
\vspace{-0.8cm}
\centering
\setlength{\abovecaptionskip}{0.cm}
\includegraphics[width=6.1in]{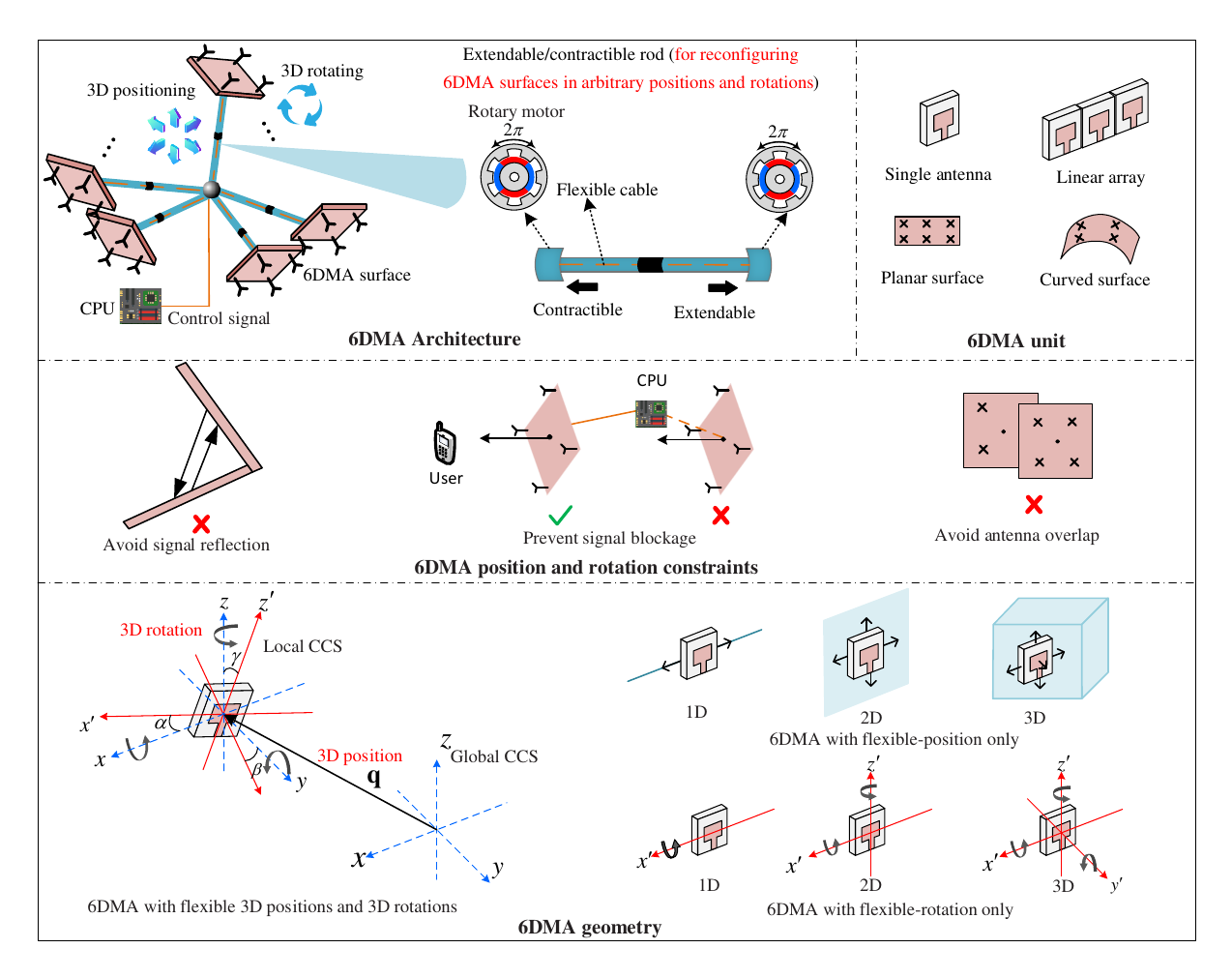}
\caption{6DMA modeling.}
\label{BS}
\vspace{-0.69cm}
\end{figure*}

As shown in Fig. \ref{BS}, the 6DMA-aided transmitter/receiver is equipped with multiple antennas/antenna surfaces. Each 6DMA surface can take various forms/shapes, such as single antenna, linear antenna array, planar  surface, curved/conformal surface, and so on.  Mathematically, the position of the $n$-th antenna of any 6DMA surface in the global Cartesian coordinate system (CCS) can be expressed as
\begin{align}\label{hu}
\mathbf{r}_{n}(\mathbf{q},\mathbf{u})=\mathbf{q}+\mathbf{R}
(\mathbf{u})\bar{\mathbf{r}}_{n},~n\in\{1,2,\cdots, N\},
\end{align}
where $N$ denotes
the total number of antennas within each 6DMA surface, $\mathbf{q}\in\mathbb{C}^{3\times 1}$ represents the 3D position of the 6DMA surface's center in the global CCS, $\bar{\mathbf{r}}_{n}$ denotes the position on the $n$-th antenna of the 6DMA surface in its local CCS, $\mathbf{u}=[\alpha,\beta,\gamma]^T$ denotes the 3D
rotation of the 6DMA surface with $\alpha\in[0,2\pi)$, $\beta\in[0,2\pi)$, and $\gamma\in[0,2\pi)$ representing the rotation angles in different directions (see Fig. \ref{BS}), and $\mathbf{R}(\mathbf{u})$ denotes the corresponding rotation matrix \cite{shao20246d,6dma_dis}. In this model, the position and rotation of each 6DMA surface can be characterized by six parameters, i.e., the 3D position and the 3D rotation in a given 3D space at the transmitter/receiver.

General speaking, the positions and rotations
of all distributed 6DMA surfaces can be independently controlled to cater to the space-/time-varying channel distribution in wireless networks. This can be achieved by attaching motors to all 6DMA surfaces. Moreover, as shown in Fig. \ref{BS}, each
6DMA surface is connected with a central processing unit (CPU) at the transceiver via a separate rod. Specifically, two rotary motors are mounted at the two ends of the rod and controlled by the CPU to adjust the position and rotation of each 6DMA surface, respectively. In addition, each rod can flexibly contract and extend to control the distance between each 6DMA surface and the CPU. Each rod also contains flexible wires (e.g., coaxial cable) that provide power supply to the 6DMA surface as well as enable the control/radio-frequency (RF) signal exchange between it and the CPU. Thus, the transmitter/receiver can flexibly adjust the 3D positions and 3D rotations of all 6DMA surfaces to optimize the wireless network performance.

However, the following movement constraints should be considered in practice for jointly designing 6DMA surfaces' positions and rotations (see Fig. \ref{BS}):
{\textbf {\emph{1) Rotation Constraints to Avoid Signal Reflection}}}:
Rotation constraints are necessary to prevent mutual signal reflections between any two 6DMA surfaces. To achieve this, each 6DMA surface should not form an acute angle with any of the other 6DMA surfaces.
{\textbf {\emph{2) Rotation Constraints to Avoid Signal Blockage}}}:
Additional rotation constraints are needed to prevent each 6DMA surface from facing towards the CPU of
the transceiver which causes signal blockage. This can be achieved by tuning the normal vector of each 6DMA surface so that it does not point towards the CPU.
{\textbf {\emph{3) Minimum-Distance Constraint}}}:
A minimum distance between the centers of any pair of 6DMA surfaces is also necessary to avoid their overlap and mutual coupling. As will be discussed later, the above practical constraints render the joint position and rotation optimization of 6DMA surfaces a more challenging problem to solve.
\subsection{6DMA Channel Model}
In general, the effective channel between a pair of 6DMA-based transmitter and receiver is determined by three main components, namely, the 6D steering vector of each 6DMA surface (array), the radiation pattern of individual antennas on each 6DMA surface, and the propagation channel between the transmitter and receiver (see Fig. \ref{BS}) \cite{shao20246d, 6dma_dis}.
The 6DMA channel differs from that of conventional FPAs as it depends on not only the location of the transmitter/receiver, but also the 3D positions as well as 3D rotations of 6DMA surfaces. Therefore, by adjusting the positions and/or rotations of 6DMA surfaces, the transmitter/receiver can effectively improve the channel conditions to enhance the performance of wireless systems, as will be elaborated in the next section.

\section{6DMA Performance Enhancement}
In this section, we discuss the main performance enhancement brought by  6DMA for wireless communication/sensing, in terms
of array gain, spatial multiplexing gain, interference mitigation, and geometric gain of MIMO systems.
\subsection{Array Gain}
6DMA system can achieve a higher array gain than FPA arrays by positioning and rotating the array towards the desired direction, thus maximally exploiting the directive radiation pattern of all the antenna elements on each 6DMA surface. As shown in Fig. 2, as practical wireless systems usually adopt directional antennas, aligning the position/rotation of each 6DMA surface with the desired signal's direction can help compensate for its path loss effectively. This is particularly useful in rich-scattering propagation environments with severe multi-path fading. For example, tuning the 6DMA surface's position and rotation to align with the dominant line-of-sight (LOS) channel path can significantly enhance its strength with respect to all the other non-LOS (NLOS) channel paths, thus improving the average signal power as well as decreasing the outage probability at the receiver.

\subsection{Spatial Multiplexing Gain}
By employing 6DMAs at the transmitter and/or
receiver, the spatial multiplexing gain of MIMO systems can be improved.
Conventional MIMO systems usually cannot achieve the full spatial multiplexing gain for rate maximization due to the MIMO channel rank deficiency problem caused by various factors such as the antennas' mutual coupling and insufficient multi-path scattering.
In contrast, 6DMA surfaces can be optimally positioned/rotated based on the propagation channel between the MIMO transmitter and receiver to achieve more balanced singular values of the MIMO channel matrix, thus boosting the MIMO system capacity. The spatial multiplexing gain of 6DMA can be obtained in practice for both the point-to-multipoint (multiuser MIMO) and point-to-point (single-user MIMO) systems by designing the positions/rotations of 6DMA surfaces based on the spatial user/scatterer distribution, respectively (see Fig. 2).

\subsection{Interference Suppression}
In multi-user scenarios, 6DMAs can not only enhance the signal power in desired directions, but also suppress the interference to/from undesired directions, as shown in Fig. \ref{gain}. For example, with 6DMA surfaces' positions and rotations tuned at the BS based on the users' (long-term) channel spatial distribution, the transmit/receive beamforming at the BS based on their instantaneous channels will become more effective for interference suppression.
\subsection{Geometric Gain}
\begin{figure}[t!]
\vspace{-0.6cm}
\centering
\setlength{\abovecaptionskip}{0.cm}
\includegraphics[width=3.5in]{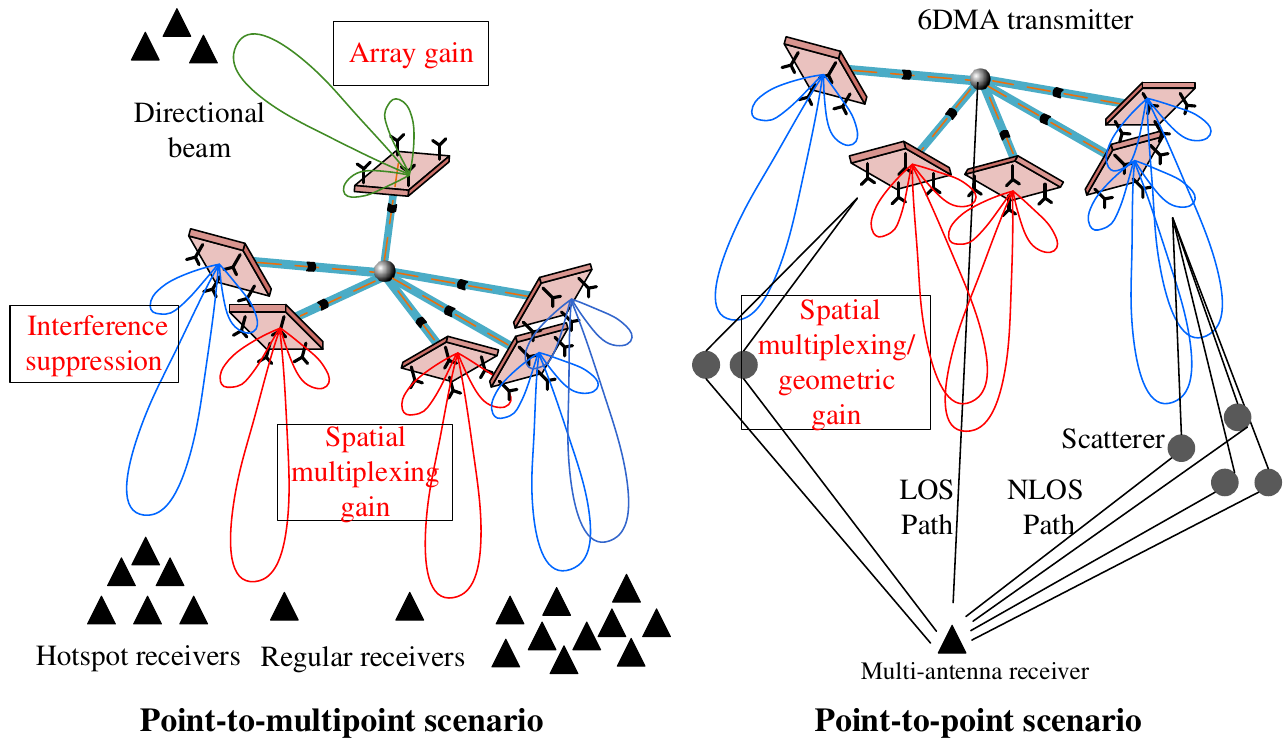}
\caption{6DMA performance enhancement.}
\label{gain}
\vspace{-0.69cm}
\end{figure}

In addition to enhancing the wireless communication performance, 6DMA can
help improve the accuracy of wireless sensing by better exploiting the inherent geometry flexibility of distributed and position/rotation-adjustable 6DMA surfaces. For example, when we consider the scatterer in Fig. \ref{gain} as the target for sensing/localization, the accuracy depends on the relative geometry between the target and the transmitter antennas, which can be optimized to yield a so-called geometric gain \cite{geo}.
Thus, how to design the 6DMA surfaces' positions and rotations to exploit the geometric gain and thereby enhance the target sensing/localization performance is an important problem to investigate in future work.

\section{6DMA Implementations}
In this section, we discuss the different ways to implement 6DMA in wireless systems, including discrete position and rotation, partially movable 6DMA, pre-configured 6DMA, and passive 6DMA.
\subsection{Discrete Position and Rotation}
While continuously tuning the position and/or rotation of each 6DMA
suffice offers the greatest flexibility and thus leads to the highest performance gains over FPAs, it is difficult
to realize in practice since 6DMA surfaces need to be mechanically moved by physical devices such as rotary motors, which can only adjust the position/rotation of each 6DMA surface in discrete steps.
Moreover, the hardware cost, power consumption, and movement time overhead increase with the number of 6DMA surfaces to cover a large movement region at the transmitter/receiver. As such, it is more cost-effective and practically viable to implement a finite number of discrete position and rotation levels for 6DMA surfaces to properly balance performance gain and implementation cost.
\begin{figure*}[t!]
\vspace{-0.8cm}
\centering
\setlength{\abovecaptionskip}{0.cm}
\includegraphics[width=6.99in]{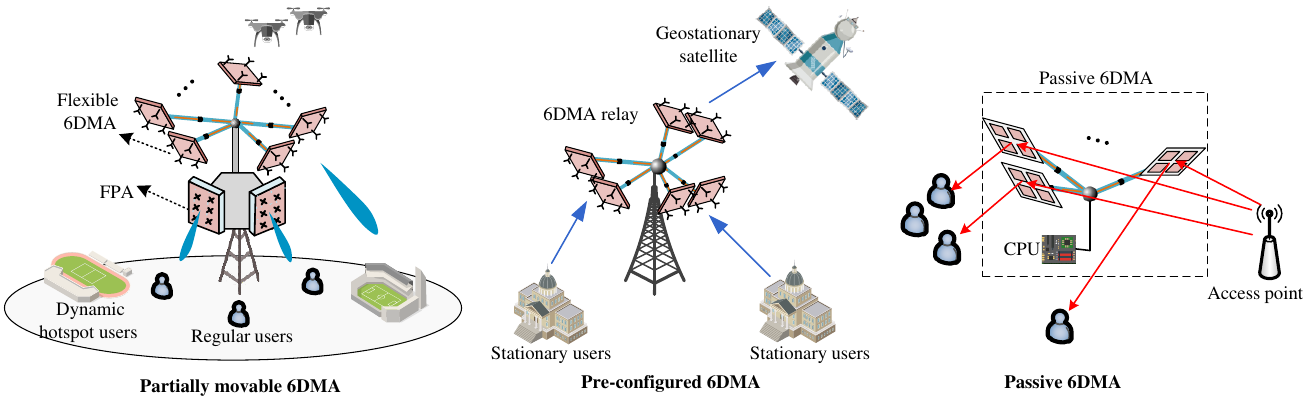}
\caption{6DMA practical implementation in wireless network.}
\label{system_model}
\vspace{-0.69cm}
\end{figure*}

\subsection{Partially Movable 6DMA}
Although freely positioning/rotating
6DMAs offers the greatest flexibility and thus highest
capacity improvement, its implementation may be challenging
in practice due to the drastic architecture change required for
the existing transmitter/receiver with the FPA array, which may lead to a
significant increase in infrastructure cost.
Alternatively, partially movable 6DMA (see Fig. \ref{system_model}) provides a good balance between performance and implementation cost, which consists of both position and rotation-adjustable
6DMAs for serving the users in hotspot areas and conventional FPA arrays for serving the remaining regular users \cite{xiaoming}. Furthermore, in scenarios with fixed transmitter and receiver locations and static environments, 6DMA is simplified to pre-configured 6DMA, where the 6DMA surfaces at the transmitter/receiver are pre-designed and do not need to move in real-time after installment. For example, a pre-configured 6DMA-enabled relay can be deployed in a remote area to help establish the communication links between distributed users at fixed locations and a Geostationary (GEO) satellite (see Fig. \ref{system_model}).
\subsection{Passive 6DMA}
Existing works on 6DMA (e.g., \cite{shao20246d, 6dma_dis}) mainly consider 6DMA surfaces deployed at the transmitter/receiver/relay, which are  composed of active antenna elements, thus termed as active 6DMA. Alternatively, 6DMA surfaces can also be made of passive reflecting elements (like IRS/RIS) and deployed in the environment to smartly reconfigure the wireless signal propagation (see Fig. \ref{system_model}). Such passive 6DMA is more cost-effective and energy-efficient as compared to active 6DMA and thus can be densely deployed in the wireless network.
By optimizing the positions and rotations of passive 6DMA surfaces  based on the spatial channel distribution of their assisted BS/access point and users, the wireless system performance can be enhanced over the traditional IRS/RIS \cite{proc} with fixed-position reflecting/reconfigurable elements.

\section{6DMA Applications}
In this section, we illustrate some typical applications of 6DMA in wireless networks, which are depicted in Fig. \ref{application}.
\subsection{6DMA Multiple Access}
6DMA can be used in multiple access to enhance the multi-user
system capacity without increasing the number of antennas.
As shown in Fig. \ref{application}, a 6DMA-enabled BS serves spatially distributed users in a single cell. Although the number of active users and their locations may change over time, the users' spatial channel distribution remains approximately constant over different subareas in the cell for a sufficiently long duration (e.g., hours or even longer). As a result, by deploying 6DMAs at the BS and adaptively adjusting their positions and rotations based on the non-uniform spatial distribution of the users in the cell, the network capacity can be significantly improved as compared to existing BSs with FPAs such as the three-sector antenna array \cite{shao20246d, 6dma_dis}.

\subsection{6DMA Cell-Free Network}
It is also practically appealing to apply 6DMA in the cell-free network to improve its performance. As shown in Fig. \ref{application}, distributed 6DMA BSs are deployed in a given area to cooperatively serve the users therein via joint signal transmission/reception as well as joint antenna position/rotation optimization. Moreover, for ease of implementation, a cell-free network in practice usually groups the distributed BSs into different clusters with intra-cluster joint signal processing, while their inter-cluster interference may still exist which needs to be properly controlled. In this case, deploying 6DMA BS at the boundary between adjacent clusters provides an effective means for interference mitigation by appropriately adjusting the positions and rotations of its 6DMA surfaces towards the desired cluster direction.

\subsection{6DMA for UAV}
6DMA can also be mounted on
unmanned aerial vehicles (UAVs) to enhance their communication/sensing performance. In conventional UAV communication/sensing systems, a main issue is the blockage of the LOS channel between the UAV and the communication user/sensing target on the ground \cite{uavv}. Although this issue can be overcome by designing the UAV placement/trajectory to avoid the LOS link blockage \cite{uavv}, it requires UAV movement which incurs additional power consumption and time delay. In contrast, by deploying 6DMA surfaces at the UAV and adjusting their positions and rotations based on the locations of the communication users/sensing targets, the establishment of LoS links becomes easier, with less UAV movement required (see Fig. \ref{application}).

\subsection{6DMA for User Terminal}
In addition to equipping 6DMA at the BS/relay, 6DMAs can be installed at the user terminal to enhance its communication/sensing performance, provided that there is sufficient space for their positioning/rotation. In this regard, two applicable scenarios are 6DMA-enabled ground terminals in low Earth orbit (LEO) satellite communication system \cite{LEO} and 6DMA-enabled vehicles in vehicular communication network \cite{vec}. As shown in Fig. \ref{application}, by taking the vehicular network as an example, due to frequent handovers of high-speed vehicles with different serving BSs, the communication interruption caused by handover failures can severely degrade the communication experience of the users inside the vehicles. To enhance handover success rate and ensure seamless coverage, mounting 6DMAs on top of the vehicles allows for the dynamic allocation of antenna resources communicating with different serving BSs concurrently by adjusting 6DMAs' positions and rotations, thus improving the communication rate and reliability of high-speed vehicular users.

\subsection{6DMA for Sensing and ISAC}
\begin{figure*}[t!]
\vspace{-0.8cm}
\centering
\setlength{\abovecaptionskip}{0.cm}
\includegraphics[width=6.9in]{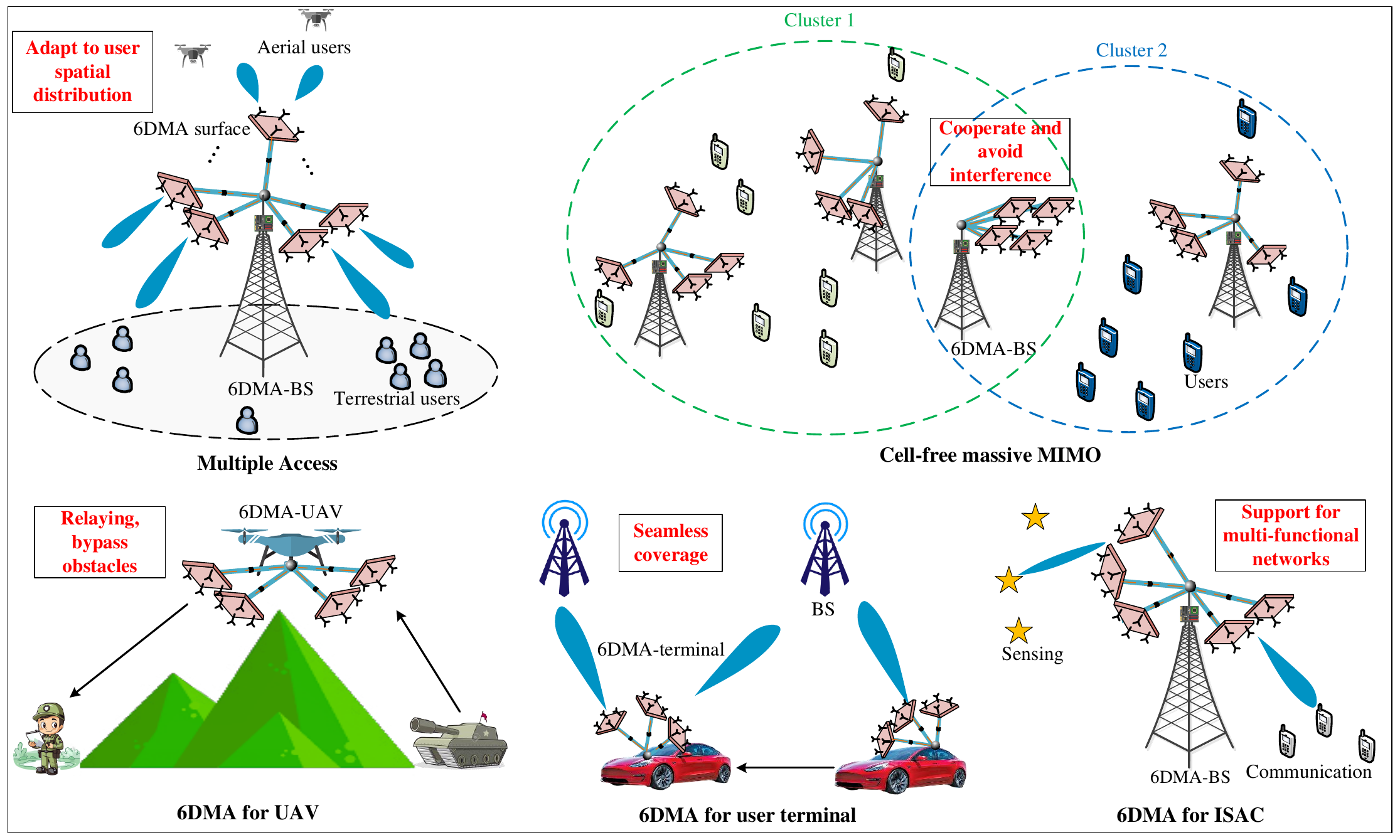}
\caption{Typical 6DMA applications in wireless network.}
\label{application}
\vspace{-0.69cm}
\end{figure*}

As 6DMA is promising to help improve the performance of wireless sensing (see Section III-D), it is also envisioned as a potential enabling technology for the  integrated sensing and communications (ISAC) in future wireless networks \cite{isac}. By sharing the network resources for both communications and sensing, ISAC is a cost-effective approach for realizing multi-functional wireless networks by exploiting their joint benefits and avoiding mutual interference. This can be efficiently achieved by properly allocating the 6DMAs based on the communication/sensing requirement and the user/target spatial channel distribution (see Fig. \ref{application}).

\section{Challenges and Promising Solutions}
In this section, we discuss the main challenges in designing 6DMA-aided wireless systems and present some promising approaches to tackle them.
\subsection{6DMA Position and Rotation Optimization}
One challenge of designing the 6DMAs' position and rotation lies in the practical position and rotation constraints as discussed in Section II-A. These constraints are non-convex in general and further complicated by the coupling between position and rotation variables. Thus, it is difficult to find the optimal 6DMAs' positions and rotations subject to these constraints which yield the best network performance. A practical approach to tackle this challenge is by applying the
alternating optimization technique to iteratively optimize the position/rotation of each 6DMA surface by fixing those
of all the others in each iteration. While in each iteration, efficient optimization methods can be used to convert non-convex position/rotation constraints into convex ones which are easier to handle. For instance, a linear approximation of the rotation matrix was applied for small rotation angle changes \cite{shao20246d}.

Another challenge in designing the position and rotation for 6DMAs in practice is due to the discrete position/rotation levels of each 6DMA surface, which leads to a very large number of position/rotation combinations over all 6DMA surfaces to search for the optimal solution.
A practical
approach is to first relax the discrete position/rotation variables into continuous ones and solve the relaxed problem using efficient optimization techniques. The discrete solutions are then obtained by e.g., quantizing the obtained solutions to their nearest values in the discrete sets \cite{6dma_dis}.
While this approach significantly reduces the computational complexity compared to exhaustive search, it may suffer various losses in performance due to quantization errors, depending on
the number of discrete position/rotation levels and the number of 6DMA surfaces. Thus, future research is needed to develop more efficient methods for high-performance discrete solutions under practical constraints.

Based on the type of channel state information (CSI) and its availability, 6DMA position and rotation optimization can be classified into three categories, explained as follows. {\textbf {\emph{1) Instantaneous CSI-based approach}}}: In a quasi-static environment with slowly-varying channels, 6DMAs' position and rotation can be adapted based on the instantaneous CSI to achieve higher performance gains over FAS/MA with antenna position adaptation only. However, this approach is difficult to implement in fast fading channels due to the need for frequent and high-speed movement of the antennas.
{\textbf {\emph{2) Statistical CSI-based approach}}}: To avoid the frequent movement of 6DMAs based on the instantaneous CSI, the position and rotation of 6DMAs can be designed according to the statistical CSI (e.g., user spatial distribution, channel statistical covariance matrix) which changes much more slowly as compared to instantaneous CSI. With the knowledge of statistical CSI, instantaneous CSI can be generated accordingly offline via Monte Carlo simulations, based on which the 6DMA's position and rotation can be designed to maximize the average capacity in the long term \cite{shao20246d, 6dma_dis}. {\textbf {\emph{3) No prior CSI}}}: For the case without any prior knowledge of CSI, the 6DMA's position and rotation can be optimized based on the real-time training data measured in terms of the performance metric of interest (e.g., signal-to-noise ratio (SNR), rate) with different combinations of training positions and rotations, by applying e.g., the conditional sample mean (CSM) algorithm \cite{6dma_dis}. In the future, more research endeavors are needed to further investigate the 6DMA's position and rotation optimization under different CSI availability cases to meet different system requirements under different practical setups.

Finally, it is worth noting that the position and rotation optimization
of 6DMAs needs to take into account the antenna radiation pattern and other wireless system design considerations such as power control, beamforming, and user transmission scheduling to achieve optimal performance. As these considerations will have an impact on the 6DMA's position and rotation optimization and its performance gain, their joint design is important and worth perusing in future work.

\subsection{6DMA Channel Estimation}
For 6DMA systems, channel estimation aims to acquire the instantaneous/statistical CSI for all available 6DMA positions and rotations within a specified transmitter/receiver region to facilitate the 6DMA's position and rotation optimization.
Different from the channel estimation in conventional MIMO systems with a finite number of channels for a fixed number of antenna positions, the channel estimation for 6DMA systems usually involves a much larger number of channels as the 6DMAs can move to arbitrary positions/rotations in a specified region. Even for the case of discrete 6DMA positions and rotations, it is practically difficult to measure the channels at all the possible positions and rotations by physically moving the 6DMA to them, which incurs prohibitively high channel training overhead. To reduce such training overhead, sparse signal recovery algorithms, such as compressed sensing, can be employed to estimate instantaneous CSI for 6DMA systems based on channel measurements at a small set of antenna positions/rotations by exploiting the channel multi-path sparsity in the angular domain.
However, the channel sparsity assumption does not always hold in practice and the resulting channel model mismatch may lead to unpredictable performance loss of 6DMA channel estimation. Therefore, non-parametric channel estimation methods, e.g., successive Bayesian reconstruction and deep learning are promising and worthy of investigation in future work. To further reduce the training overhead, statistical CSI estimation
is practically appealing as it does not change frequently and there are rich statistical signal processing techniques such as  maximum-likelihood (ML) estimation and expectation-maximization (EM) estimation to efficiently attain the statistical CSI.

\subsection{Movement Scheduling}
After obtaining the optimized positions and rotations of 6DMAs, another practical problem that needs to be solved is movement scheduling, which deals with how to move distributed 6DMA surfaces over time to their desired positions/rotations. Since moving 6DMA surfaces to their new positions/rotations causes temporary channel changes which may interrupt the current communication performance, the movement scheduling of 6DMAs at all transmitters/receivers needs to be jointly designed to minimize the performance loss as well as the movement energy and time cost. Promising approaches to movement scheduling for 6DMA surfaces include parallel/sequential movement, collaborative movement at different transmitters/receivers, etc., which require further investigation in future work.
\section{Numerical results}
\begin{figure}
\vspace{-0.7cm}
\centering
\includegraphics[width=2.9in]{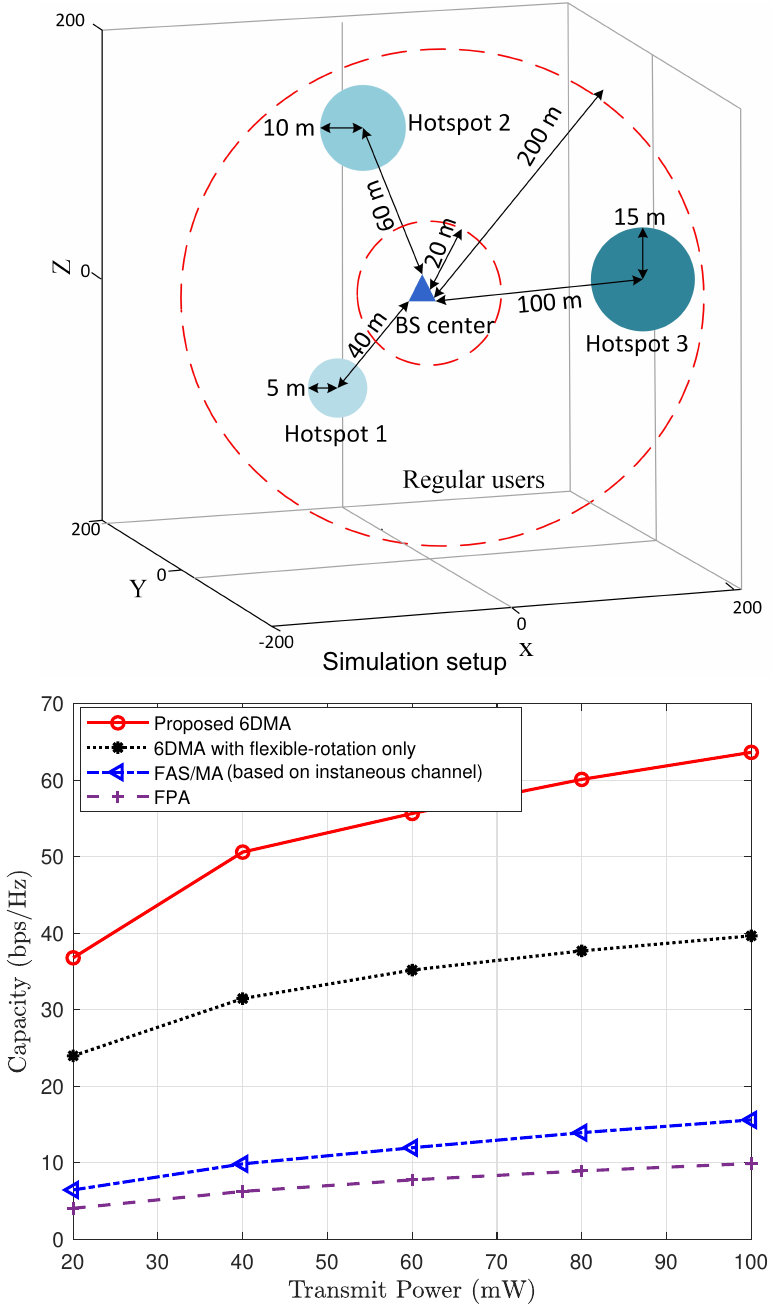}
      \caption{6DMA simulation setup and capacity improvement.}
\label{powerstep}
\vspace{-0.69cm}
\end{figure}

We consider an uplink communication system, where a BS is equipped with $16$ 6DMA surfaces with $4$ antennas on each surface. The BS serves multiple single-FPA users whose locations are depicted in Fig. \ref{powerstep}. The average numbers of users in the three hotspot sub-regions 1, 2, and 3 follow the ratio of 1:2:3, while the remaining region accommodates regular users. The directive antenna pattern is set according to the standard of 3GPP.
The positions and rotations of 6DMA surfaces are optimized based on the users' statistical CSI \cite{shao20246d}.
We compare the proposed 6DMA system with the following three schemes, all of which are based on the BS with three-sector antenna arrays, and the total number of antennas are the same for all schemes. 1) FPA: The 3D locations and 3D rotations of all sector antenna arrays are fixed. 2) 6DMA with flexible rotation only: The
centroid of each sector antenna array is fixed,
and we apply the algorithm in \cite{shao20246d} to optimize the
rotation of each sector antenna array based on the statistical CSI. 3) FAS/MA: The rotations of all sector antenna arrays are fixed, and we apply the algorithm in \cite{shao20246d} to optimize the positions of all antennas on each sector antenna array based on the instantaneous CSI (as FAS/MA mainly exploit the small-scale channel fading).

In Fig. \ref{powerstep}, we show the network capacity of
the proposed and benchmark schemes versus user transmit power. It is observed that with the same transmit power, the 6DMA system can achieve larger network capacity as compared to the FPA and FAS/MA schemes.
This is because the 6DMA scheme has higher spatial DoFs and can
deploy the antenna resources more reasonably to match the
channel spatial distribution. In contrast, the FAS/MA scheme can only adjust the antenna positions within a 2D surface to mitigate the impact of deep fading. Furthermore, it is shown that the performance gap increases as the transmit power increases. This is expected as the network capacity becomes more interference-limited as the transmit power or received SNR increases.

\section{Conclusions}
In this article, we provide an overview of the
promising 6DMA technology for enhancing wireless network performance by fully exploiting the position and rotation adjustment of distributed antennas at the transmitter/receiver. As
6DMA-aided wireless networks are new and remain
largely unexplored, it is hoped that this article
would provide a useful guide for
the future research on them.
In particular, we
foresee that the integration of 6DMAs into future
wireless networks will fundamentally improve antenna agility and adaptability, and bring fertile new opportunities for applications and research.
\bibliographystyle{IEEEtran}
\bibliography{fabs}

\section*{Biographies}
\vspace{-253pt}
\begin{IEEEbiographynophoto}{Xiaodan Shao}
[M'22] (xiaodan.shao@fau.de) is a Humboldt Research Fellow with the Institute for Digital Communications, Friedrich-Alexander-University Erlangen-Nuremberg (FAU), Erlangen, Germany.
\end{IEEEbiographynophoto}
\vspace{-253pt}
\begin{IEEEbiographynophoto}{Rui Zhang}
[F'17] (elezhang@nus.edu.sg) is a Professor with School of Science and Engineering, Shenzhen Research Institute of Big Data, The Chinese University of Hong Kong, Shenzhen, China, ECE Department of National University of Singapore, Singapore.
\end{IEEEbiographynophoto}
\end{document}